\documentclass{emulateapj}
\usepackage{verbatim}
\usepackage{apjfonts}

\slugcomment{Accepted for publication in ApJ Letters}

\shorttitle{$^{13}$CO $J = 6 \rightarrow 5$ in NGC 253}
\shortauthors{Hailey-Dunsheath et al.}

\begin{document}


\title{Detection of the $^{13}$CO $J = 6 \rightarrow 5$ transition in the Starburst Galaxy NGC 253}





\author{S. Hailey-Dunsheath\altaffilmark{1}, T. Nikola\altaffilmark{1}, G. J. Stacey\altaffilmark{1}, T. E. Oberst\altaffilmark{1,2}, S. C. Parshley\altaffilmark{1}, C. M. Bradford\altaffilmark{3}, P. A. R. Ade\altaffilmark{4}, and C. E. Tucker\altaffilmark{4}}


\altaffiltext{1}{Department of Astronomy, Cornell University, Ithaca, NY 14853; steve@astro.cornell.edu.}
\altaffiltext{2}{Current Address: Westminster College, Department of Physics, 319 S. Market St., New Wilmington, PA 16172.}
\altaffiltext{3}{Jet Propulsion Laboratory, Pasadena, CA 91109.}
\altaffiltext{4}{Department of Physics and Astronomy, Cardiff University, Cardiff CF24 3AA, UK.}


\begin{abstract}
We report the detection of $^{13}$CO $J = 6 \rightarrow 5$ emission from the nucleus of the starburst galaxy NGC 253 with the redshift (z) and Early Universe Spectrometer (ZEUS), a new submillimeter grating spectrometer. This is the first extragalactic detection of the $^{13}$CO $J = 6 \rightarrow 5$ transition, which traces warm, dense molecular gas. We employ a multi-line LVG analysis and find $\approx$ $35\% - 60\%$ of the molecular ISM is both warm ($T \sim 110$ K) and dense ($n_\mathrm{H_\mathrm{2}} \sim 10^4$ cm$^{-3}$). We analyze the potential heat sources, and conclude that UV and X-ray photons are unlikely to be energetically important. Instead, the molecular gas is most likely heated by an elevated density of cosmic rays or by the decay of supersonic turbulence through shocks. If the cosmic rays and turbulence are created by stellar feedback within the starburst, then our analysis suggests the starburst may be self-limiting.
\end{abstract}




\defcitealias{Bradford2003}{B03}
\section{Introduction}

NGC 253 is a nearby \citep[d $\approx$ 2.5 Mpc;][]{Mauersberger1996}, highly inclined \citep[i $\approx$ 78$\degr$;][]{Pence1981} Sc galaxy undergoing a nuclear starburst \citep{Rieke1980}. The IR luminosity in the central 30$\arcsec$ is 1.5 $\times$ 10$^{10}$ L$_{\sun}$ \citep{Telesco1980}, and the mid-IR morphology indicates most of this emission arises from a 7$\arcsec$ region centered within 1$\arcsec$ of the 2 $\mu$m nucleus \citep{Telesco1993}. Near-IR images show a prominent stellar bar \citep{Scoville1985} which likely plays a major role in channeling gas into the central starburst \citep{Peng1996,Das2001}. Estimates of the gas mass in the central $\sim$ 300 pc range from $0.4 - 4.2$ $\times$ 10$^8$ M$_{\sun}$ \citep{Krugel1990,Mauersberger1996,Harrison1999}.

The $J = 6 \rightarrow 5$ and $J = 7 \rightarrow 6$ transitions of CO arise from states with energy levels 116 K and 155 K above ground, and are thus sensitive probes of the warm molecular gas found in regions of massive star formation. Observations of these lines in the starburst nucleus of NGC 253 \citep{Harris1991, Bradford2003, Bayet2004, Guesten2006} have indeed shown that much of the molecular gas is highly excited, although a consensus on the details of the excitation have yet to be reached \citep{Guesten2006}. Using a multi-line excitation analysis, \citet[hereafter B03]{Bradford2003} find the central 180 pc contain a large mass ($2 - 5$ $\times$ 10$^7$ M$_{\sun}$) of warm (T $\sim$ 120 K), dense ($n_{\mathrm{H_\mathrm{2}}} \sim 4.5 \times 10^{4}$ cm$^{-3}$) molecular gas, most likely heated by cosmic rays injected into the ISM by the many supernovae \citep[$\sim 0.1$ yr$^{-1}$;][]{Ulvestad1997}. This model finds large optical depths in the mid-J $^{12}$CO lines, and thus predicts bright $^{13}$CO emission. To further constrain the excitation and energetics of the molecular gas we observed the $^{13}$CO $J = 6 \rightarrow 5$ transition, which provides a strong constraint on the $^{12}$CO $J = 6 \rightarrow 5$ opacity. This is the first extragalactic detection of the $^{13}$CO $J = 6 \rightarrow 5$ transition, and the first detection of any $^{13}$CO transition greater than $J = 3 \rightarrow 2$ from beyond the Magellanic Clouds.

\section{Observations}

We observed $^{12}$CO $J = 6 \rightarrow 5$ (433.56 $\mu$m), $J = 7 \rightarrow 6$ (371.65 $\mu$m), $^{13}$CO $J = 6 \rightarrow 5$ (453.50 $\mu$m), and the [CI] $^3$P$_2$ $\rightarrow$ $^3$P$_1$ fine-structure line (370.41 $\mu$m) toward NGC 253 in December 2006 with ZEUS \citep{Stacey2004} at the Caltech Submillimeter Observatory (CSO) on Mauna Kea, Hawaii. ZEUS is a direct-detection grating spectrometer providing a slit-limited resolving power of $\lambda$/$\Delta\lambda$ $\sim$ 1000 across the 350 $\mu$m and 450 $\mu$m telluric windows. It currently utilizes a 1 $\times$ 32 semiconductor bolometer array oriented along the dispersion direction, with the pixel size approximately matched to a spectral resolution element. A pair of bandpass filters centered at 350 $\mu$m and 450 $\mu$m are mounted directly in front of the detector array, such that the system simultaneously provides a 16 pixel spectrum in both windows. The bandwidth is sufficiently large to simultaneously observe $^{12}$CO $J = 7 \rightarrow 6$ and [CI].

We obtained absolute spectral calibration with observations of Orion (BN-KL), and flux calibration with observations of Saturn, which was assumed to have brightness temperatures of 116 K, 118 K, and 97 K at 434 $\mu$m, 453 $\mu$m, and 371 $\mu$m, respectively \citep{Hildebrand1985,Orton2000}. We observed the four lines over the course of three nights in good submillimeter weather, with $\tau_{225\,}$$_\mathrm{GHz}$ $= 0.04 - 0.06$. A zenith opacity was obtained from both $\tau_{225\,}$$_\mathrm{GHz}$ and $\tau_{350\,}$$_\mathrm{\mu m}$ using the CSO atmospheric transmission model, and the mean of the two values was used to calculate the transmission to the source. Observations of NGC 253 were centered at R.A. = 00$^\mathrm{h}$47$^\mathrm{m}$33$\fs$2, decl. = $-$25$\arcdeg$17$\arcmin$18$\arcsec$ (J2000.0), and small maps in the $^{12}$CO lines verified that the beam was centered within 4$\arcsec$ of the CO emission peak. We used total power maps of Uranus to measure the FWHM of the beam to be 11$\arcsec$ at 434 $\mu$m and 453 $\mu$m and 10$\arcsec$ at 371 $\mu$m. All data were obtained by chopping and nodding the telescope with a 30$\arcsec$ throw. The spectra of $^{12}$CO $J = 6 \rightarrow 5$, $^{13}$CO $J = 6 \rightarrow 5$, and the $^{12}$CO $J = 7 \rightarrow 6$ and [CI] pair shown in Figure 1 represent total integrations times of 6, 70, and 5 minutes, respectively. A linear baseline is removed from all spectra, and the integrated intensities are listed in Table 1. In addition to the nuclear spectra we obtained a simultaneous map in the $^{12}$CO $J = 7 \rightarrow 6$ and [CI] lines, which will be presented elsewhere (T. Nikola et al. 2008, in preparation).
\begin{figure}
\plotone{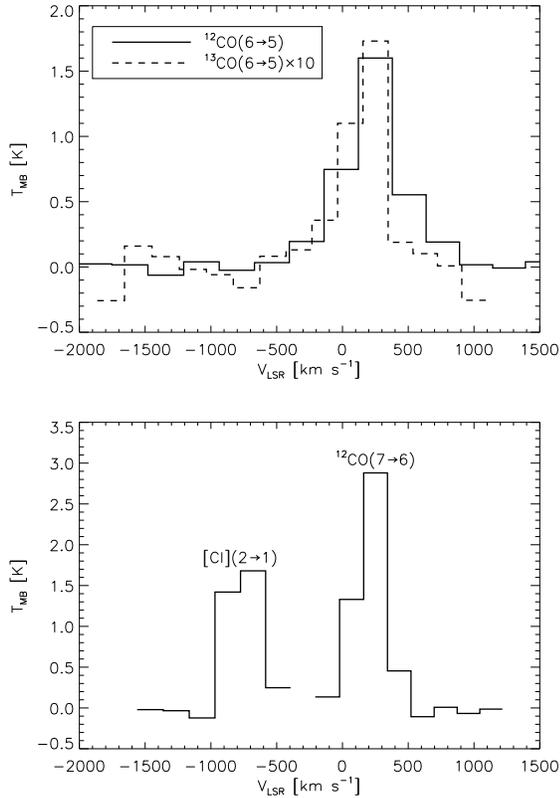}
\caption{\textit{Top:} Spectra of $^{12}$CO $J = 6 \rightarrow 5$ and $^{13}$CO $J = 6 \rightarrow 5$ (scaled by $\times$10). \textit{Bottom:} Spectrum of $^{12}$CO $J = 7 \rightarrow 6$ and [CI] $^3$P$_2$ $\rightarrow$ $^3$P$_1$ with a rogue pixel removed near the center. The velocity scale is referenced to $^{12}$CO $J = 7 \rightarrow 6$.\label{fig1}}
\end{figure}

\section{Results}
\subsection{LVG Model} \label{lvgmodel}

To examine the CO excitation we assemble the lower-J line intensities from the literature, and correct all measurements to a common 15$\arcsec$ beam. The $^{12}$CO $J = 4 \rightarrow 3$ intensity is obtained from \citet{Guesten2006}, and the $J = 3 \rightarrow 2$ and lower transitions are taken from \citet{Harrison1999}. The $^{12}$CO $J = 6 \rightarrow 5$ map of \citet{Bayet2004} is used to correct the intensities measured here to a 15$\arcsec$ scale, and when necessary the intensities obtained from \citet{Harrison1999} are corrected using power law interpolations as outlined in \citetalias{Bradford2003}. The line intensities used in our analysis are listed in Table 1.

To quantitatively analyze the CO line SED we employ a large velocity gradient (LVG) model, in which the excitation and opacity of the CO are determined by a gas density ($n_{\mathrm{H_\mathrm{2}}}$), kinetic temperature ($T_\mathrm{kin}$), and CO abundance per velocity gradient ([CO/H$_2$]/$dv/dr$). We use an escape probability formalism with $\beta = (1-e^{-\tau})/\tau$, derived for a spherical cloud undergoing uniform collapse \citep{Castor1970,Goldreich1974}. The source is assumed to contain a large number of these unresolved clouds, such that the absolute line intensities are proportional to a beam-averaged CO column density ($N_{\mathrm{CO}}$). We increase the CO-H$_2$ collisional rate coefficients from \citet{Flower2001} by 21\% to account for collisions with He \citepalias[][and references therein]{Bradford2003}, and fix the H$_2$ ortho/para ratio at 3. The CO abundance is set to [CO/H$_2$] = 8.5 $\times$ 10$^{-5}$ \citep{Frerking1982} and the isotopologue abundance ratio to [$^{12}$CO/$^{13}$CO] = 40 \citep{Henkel1993}.
\begin{table}
\begin{center}
\caption{Integrated CO Line Intensities\label{tbl-3}}
\begin{tabular}{cccccrrrrr}
\tableline\tableline
Transition                   & Beam        & I                    & I                                    & $\sigma$ \\    
                             & [$\arcsec$] & [K km s$^{-1}$]      & [ergs s$^{-1}$ cm$^{-2}$ sr$^{-1}$]   & [\%]     \\    
\tableline
$^{12}$CO(6$\rightarrow$5)   &11           & 854 $\pm$ 20         &2.89 $\times$ 10$^{-4}$               &30        \\
                             &15           & 573                  &1.94 $\times$ 10$^{-4}$               &30        \\
$^{12}$CO(7$\rightarrow$6)   &10           & 694 $\pm$ 30         &3.73 $\times$ 10$^{-4}$               &30        \\
                             &15           & 418\tablenotemark{a}\tablenotemark{,b} &2.25 $\times$ 10$^{-4}$               &30        \\
$^{13}$CO(6$\rightarrow$5)   &11           & 65 $\pm$ 5           &1.97 $\times$ 10$^{-5}$               &30        \\
                             &15           & 43                   &1.29 $\times$ 10$^{-5}$               &30        \\
\tableline
$^{12}$CO(1$\rightarrow$0)   &15           & 1105                 &1.73 $\times$ 10$^{-6}$               &22        \\
$^{12}$CO(2$\rightarrow$1)   &15           & 1500                 &1.88 $\times$ 10$^{-5}$               &16        \\
$^{12}$CO(3$\rightarrow$2)   &15           & 1134                 &4.81 $\times$ 10$^{-5}$               &14        \\
$^{12}$CO(4$\rightarrow$3)   &15           & 1040                 &1.04 $\times$ 10$^{-4}$               &15        \\
$^{13}$CO(1$\rightarrow$0)   &15           & 94                   &1.29 $\times$ 10$^{-7}$               &20        \\
$^{13}$CO(2$\rightarrow$1)   &15           & 113                  &1.24 $\times$ 10$^{-6}$               &12        \\
$^{13}$CO(3$\rightarrow$2)   &15           & 117                  &4.33 $\times$ 10$^{-6}$               &14        \\
\tableline
\end{tabular}
\tablecomments{Intensities of $^{12}$CO $J = 6 \rightarrow 5$, $J = 7 \rightarrow 6$, and $^{13}$CO $J = 6 \rightarrow 5$ from this work, and lower-J intensities from \citet{Guesten2006} and \citet{Harrison1999}. All measurements have been corrected to 15$\arcsec$ using a methodology described in \S~\ref{lvgmodel}. Intensities measured here have statistical errors listed in column 3, and total uncertainties (30\%) dominated by systematic uncertainties in the high frequency sky transmission.}
\tablenotetext{a}{10\% larger than the intensity reported by \citet{Guesten2006}, and well within the calibration uncertainties.}
\tablenotetext{b}{\citetalias{Bradford2003} report an intensity calculated for a source which couples to half the power in the Gaussian main beam, rather than to the full main beam as is done here. Reducing their value by the corresponding factor of 2 yields an intensity 10\% larger than measured here, well within the calibration uncertainties.}
\end{center}
\end{table}

\subsection{High Excitation Component} \label{highexcitation}

As for \citetalias{Bradford2003}, we find that any single set of LVG model parameters capable of producing the mid-J emission underpredicts the $J = 2 \rightarrow 1$ and $J = 1 \rightarrow 0$ intensities, necessitating the adoption of a two component model. \citet{Guesten2006} find that a single component can produce the $^{12}$CO intensities and the $J = 3 \rightarrow 2$ and lower transitions of $^{13}$CO, but such a model would not account for the bright $^{13}$CO $J = 6 \rightarrow 5$ emission measured here. We begin by using the $J = 3 \rightarrow 2$ and higher transitions to constrain the high excitation component, and then introduce a low excitation component to account for the excess $J = 2 \rightarrow 1$ and $J = 1 \rightarrow 0$ emission.

\begin{figure}
\plotone{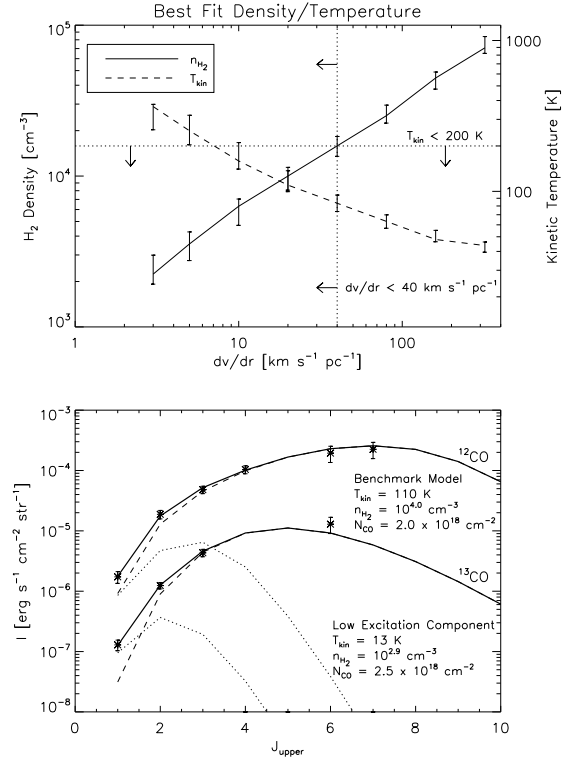}
\caption{Results of LVG analysis. \textit{Top:} Best fit values of $n_{\mathrm{H_\mathrm{2}}}$ (\textit{solid}) and $T_\mathrm{kin}$ (\textit{dashed}) as a function of $dv/dr$, with error bars showing $\pm$1$\sigma$ range of the posterior probability density functions. The dotted lines show the upper limits to $T_\mathrm{kin}$ and $dv/dr$. \textit{Bottom:} Integrated line intensities along with the benchmark high excitation model (\textit{dashed}), a representative low excitation component (\textit{dotted}), and the sum (\textit{solid}).\label{lvgresults_excitation}}
\end{figure}

We calculate a four-dimensional grid of model CO line SEDs, varying $n_{\mathrm{H_\mathrm{2}}}$, $T_\mathrm{kin}$, $dv/dr$, and $N_\mathrm{CO}$ over a large volume of parameter space. Comparing these model calculations to the observed mid-J CO line intensities, we find solutions giving $\chi^2_\mathrm{reduced}$ $\lesssim$ 1 for values of $dv/dr \gtrsim 3$ km s$^{-1}$ pc$^{-1}$. In Figure~\ref{lvgresults_excitation} we plot the values of $n_{\mathrm{H_\mathrm{2}}}$ and $T_\mathrm{kin}$ giving the best fits for velocity gradients in the range $dv/dr = 3 - 320$ km s$^{-1}$ pc$^{-1}$. Over the modeled range of $dv/dr$ these values change by an order of magnitude or more, so to further restrict parameter space we must apply prior constraints to $dv/dr$ and $T_\mathrm{kin}$.

In the LVG approximation the velocity gradient is produced by large-scale systematic motion, but for a self-gravitating cloud in virial equilibrium we can approximate $dv/dr \approx 3.1$ km s$^{-1}$ pc$^{-1}$ $\sqrt{n_{\mathrm{H_\mathrm{2}}}/10^4\;\mathrm{cm}^{-3}}$ \citep{Goldsmith2001}. Allowing that $dv/dr$ may be larger due to the presence of an additional stellar mass density or a high-pressure intercloud medium \citepalias{Bradford2003}, we set an upper limit $\sim$ 10 times larger at $dv/dr \leq$ 40 km s$^{-1}$ pc$^{-1}$. $T_\mathrm{kin}$ is restricted by the results of \citet{Rigopoulou2002}, who conclude that the bulk of the warm molecular gas traced by H$_2$ rotational transitions lies at $T =$ 195 K. As the mid-J CO transitions arise from lower energy states than those producing the H$_2$ rotational lines we expect the mid-J CO emission to trace a cooler component, and therefore require $T_\mathrm{kin} \leq$ 200 K. With these two upper limits the velocity gradient is effectively restricted to $dv/dr \approx 7 - 40$ km s$^{-1}$ pc$^{-1}$, with corresponding limits to $n_{\mathrm{H_\mathrm{2}}}$ and $T_\mathrm{kin}$ (Figure~\ref{lvgresults_excitation}).

To quantify the allowed ranges of the model parameters we adopt a Bayesian formalism and calculate a posterior probability density function for each parameter \citep[cf.][]{Ward2003}. We assume a prior expectation of uniform probability per logarithmic interval for each parameter, subject to the upper limits on $dv/dr$ and $T_\mathrm{kin}$ imposed above. We find $n_{\mathrm{H_\mathrm{2}}} = 10^{3.8} - 10^{4.1}$ cm$^{-3}$, $T_\mathrm{kin} = 80 - 200$ K, and the thermal pressure is $P/k_B = 0.8 - 1.4 \times 10^6$ cm$^{-3}$ K. The beam-averaged CO column density of this warm component is well constrained to be $N_{\mathrm{CO}} = 1.7 - 2.2$ $\times$ 10$^{18}$ cm$^{-2}$, giving an associated H$_2$ mass of $M_{\mathrm{H_\mathrm{2}}} = 1.2 - 1.6$ $\times$ 10$^7$ M$_{\sun}$ in the central 180 pc. We take as our benchmark model the best fit solution obtained by fixing $dv/dr = 20$ km s$^{-1}$ pc$^{-1}$ and plot it over the data in Figure~\ref{lvgresults_excitation}.

\subsection{Low Excitation Component} \label{lowexcitation}

The residual $J = 2 \rightarrow 1$ and $J = 1 \rightarrow 0$ intensities from the benchmark model can be produced by a broad range of low excitation components with $T_\mathrm{kin} \lesssim$ 40 K and $n_{\mathrm{H_\mathrm{2}}} \sim 10^{2.4} - 10^{3.0}$ cm$^{-3}$, contributing a beam-averaged CO column density of $N_{\mathrm{CO}} = 1.5 - 3.2$ $\times$ 10$^{18}$ cm$^{-2}$. We therefore estimate the central 180 pc contain an H$_2$ mass of $M_{\mathrm{H_\mathrm{2}}} \approx 2.9$ $\times$ 10$^7$ M$_{\sun}$, $35\% - 60\%$ of which is in a warm ($T_\mathrm{kin} \sim 110$ K), dense ($n_{\mathrm{H_\mathrm{2}}} \sim 10^{4}$ cm$^{-3}$) phase.

\subsection{Comparison with Atomic Gas} \label{atomic}

\citet{Carral1994} detect 158 $\mu$m [CII] fine-structure line emission and emission from other ionized and neutral gas tracers toward the central 45$\arcsec$ of NGC 253. Based on a combined HII region and photodissociation region (PDR) model of the line and far-IR continuum flux, they estimate an atomic PDR mass of $M_{\mathrm{HI}}$ = 2.4 $\times$ 10$^6$ M$_{\sun}$. Scaling from the $^{12}$CO $J = 1 \rightarrow 0$ morphology \citep{Paglione2004} we estimate half of the PDR emission arises from the central 15$\arcsec$, giving $M_{\mathrm{HI}}$ $\approx$ 1.2 $\times$ 10$^6$ M$_{\sun}$ in the central 180 pc, a factor of $\approx$ 12 smaller than the warm molecular gas mass. PDR models generally predict comparable amounts of warm molecular and atomic components, so we conclude that the bulk of the warm molecular gas is not UV-heated gas associated with PDRs. In the next section we explore alternative mechanisms for heating the molecular gas.

\section{Discussion: What heats the gas?}
\subsection{X-Rays} \label{xdr}

Due to their smaller cross sections, X-ray photons penetrate more deeply than UV photons into clouds and heat a larger volume of the molecular gas. In this section we consider whether an X-ray Dominated Region (XDR) can produce a warm molecular gas mass significantly in excess of the warm atomic gas mass.  

\citet{Meijerink2005} present the thermal and chemical structure of four model XDRs, with combinations of low or high density ($n_{\mathrm{H}} = 10^{3.0}$, $10^{5.5}$ cm$^{-3}$) and low or high incident X-ray flux ($F_\mathrm{X} = 1.6$, $160$ ergs s$^{-1}$ cm$^{-2}$). For each model they plot the gas abundances and temperature as a function of depth into the cloud, from which we calculate the total column densities of warm C$^+$ and CO. The 158 $\mu$m [CII] transition used to trace the warm atomic component arises from a state 91 K above ground, so we include only C$^+$ warmer than 91 K. We include all CO warmer than 80 K, the minimum temperature allowed by our CO excitation analysis. The large observed ratio of $N_{\mathrm{CO}}$/$N_{\mathrm{C^+}}$ $\approx$ 1.7 (corresponding to $M_{\mathrm{H_\mathrm{2}}}$/$M_{\mathrm{HI}}$ $\approx$ 12 as discussed in \S~\ref{atomic}) can only be produced by the high density ($n_\mathrm{H} = 10^{5.5}$ cm$^{-3}$), high flux ($F_\mathrm{X} = 160$ ergs s$^{-1}$ cm$^{-2}$) model. 

The model XDRs use densities which are an order of magnitude larger or smaller than the value of $n_\mathrm{H} = 2n_{\mathrm{H_\mathrm{2}}}$ $\sim$ $10^{4.3}$ cm$^{-3}$ indicated by our LVG analysis. Interpolating between the high density, high flux model and the lower density models, we estimate an XDR with $n_\mathrm{H} = 10^{4.3}$ cm$^{-3}$ will match the observed $N_{\mathrm{CO}}$/$N_{\mathrm{C^+}}$ ratio only if $F_\mathrm{X}$ $\gtrsim$ $10$ ergs s$^{-1}$ cm$^{-2}$. However, such an XDR will produce an [OI] 63 $\mu$m/[CII] 158 $\mu$m ratio more than $\sim$ 20 times larger than observed \citep{Carral1994,Meijerink2007}, and consequently we rule out an XDR as a potential source of the mid-J CO emission.

\subsection{Cosmic Rays} \label{crays}

It is generally accepted that low energy cosmic rays control the thermal and chemical balance in the UV-shielded inner cores of Galactic molecular clouds \citep{Goldsmith1978}. \citetalias{Bradford2003} estimates that the high supernovae rate in the nucleus of NGC 253 results in a cosmic-ray ionization rate 750 times higher than in the Galactic plane, and that these cosmic rays deposit ($5-18$) $\times$ 10$^{-25}$ ergs s$^{-1}$ per H$_2$ molecule in the molecular gas. By summing the integrated intensities predicted by our benchmark LVG model over all rotational transitions, we estimate a warm molecular gas mass of $M_{\mathrm{H_\mathrm{2}}}$ $\approx$ $1.4$ $\times$ 10$^7$ M$_{\sun}$ produces a CO luminosity of $L_\mathrm{CO}$ $\approx$ 1.6 $\times$ 10$^6$ L$_{\sun}$, corresponding to a specific cooling rate of $\approx$ 6.9 $\times$ 10$^{-25}$ ergs s$^{-1}$ per H$_2$ molecule. As this cooling rate matches the heating rate estimated by \citetalias{Bradford2003}, we suggest an elevated cosmic-ray density resulting from the starburst may provide the origin of the warm molecular gas.

\subsection{Shocks} \label{shocks}

The molecular gas in the center of NGC 253 shows evidence of shock-driven chemistry, including the large gas phase abundance of silicon \citep{Carral1994,Garcia-Burillo2000}, and the general chemical similarity to shock-dominated molecular clouds in the Galactic center \citep{Martin2006}. In this section we consider whether the mid-J CO emission may arise from shock-heated gas.

The emission from C-shocks is modeled by \citet{Draine1983} and \citet{Draine1984}. In addition to the rotational transitions of CO, the dominant coolants are the H$_\mathrm{2}$ rovibrational transitions and the 63 $\micron$ [OI] fine-structure line, and we compare observations of these tracers with the model predictions in an attempt to constrain the possible shock parameters. \citet{Engelbracht1998} suggest that thermal emission from shock-heated gas in the central 15$\arcsec$ produces an H$_2$ luminosity of $L_\mathrm{H_\mathrm{2}}$ = 1.3 $\times$ 10$^6$ L$_{\sun}$, summed over all infrared rovibrational transitions. \citet{Carral1994} measure a 63 $\mu$m [OI] luminosity of $L_\mathrm{[OI]}$ = 8.8 $\times$ 10$^6$ L$_{\sun}$ in a 42$\arcsec$ beam, of which we estimate half arises from the central 15$\arcsec$. With a total CO luminosity of $L_\mathrm{CO}$ $\approx$ 1.6 $\times$ 10$^6$ L$_{\sun}$ we find that the $L_\mathrm{H_\mathrm{2}}$/$L_\mathrm{CO} \approx 1$ and $L_\mathrm{[OI]}$/$L_\mathrm{CO} \approx 3$ ratios, as well as the general shape of the CO line SED, are reproduced for a low velocity ($v_\mathrm{shock} \lesssim 8$ km s$^{-1}$) shock incident on $n_\mathrm{H} = 10^4 - 10^5$ cm$^{-3}$ gas.

The molecular gas may be heated by shocks originating in the decay of supersonic turbulence, as is the case in the central 2 pc of the Galactic center \citep{Bradford2005}. Numerical simulations of magnetohydrodynamic (MHD) turbulence by \citet{MacLow1999} show that the conversion of dynamical to thermal energy in turbulent gas produces a specific luminosity of
\begin{displaymath} 
\frac{L}{M} = 0.36 \biggl( \frac{v_\mathrm{rms}}{8\mathrm{\;km\;s^{-1}}} \biggr) ^3 \biggl( \frac{0.1\mathrm{\;pc}}{\Lambda} \biggr) \mathrm{\frac{L_{\sun}}{M_{\sun}}},
\end{displaymath}
where $v_\mathrm{rms}$ and $\Lambda$ are the characteristic velocity and length scale of the turbulence. The highest resolution molecular maps of NGC 253 utilize a $\sim$ 2$\arcsec$ beam ($\sim$ 24 pc), so structure on sub-parsec scales remains unresolved. However, molecular gas in the Galactic center shows clumping on scales down to $\sim$ 0.1 pc \citep{Bradford2005}, so we adopt $\Lambda = 0.1$ pc. For a total molecular gas mass of $M_{\mathrm{H_\mathrm{2}}}$ $\approx$ $2.9$ $\times$ 10$^7$ M$_{\sun}$ the luminosity to mass ratio observed in the primary shock coolants is $\approx$ 0.25 L$_{\sun}$/M$_{\sun}$, comparable to the value obtained from the above equation by setting $v_\mathrm{rms} = v_\mathrm{shock} = 8$ km s$^{-1}$. We conclude that in addition to an elevated density of cosmic rays, the dissipation of turbulent energy through low velocity shocks can produce the warm molecular gas emitting in mid-J CO lines. 

By keeping a large fraction of the molecular gas warm and therefore less susceptible to gravitational instability, cosmic rays and the decay of turbulence work to halt the starburst. As the cosmic rays are produced in supernovae and the turbulence may be driven by stellar feedback, we suggest the starburst in the nucleus of NGC 253 may be self-limiting.

\acknowledgments

This work was supported by NSF grants AST-0096881, AST-0352855, AST-0705256, and AST-0722220, and by NASA grants NGT5-50470 and NNG05GK70H. We are indebted to the GSFC group (C. A. Allen, S. H. Moseley, D. J. Benford, and J. G. Staguhn) for their sensitive bolometers. We also thank the CSO staff for their support of ZEUS operations, and an anonymous referee for many helpful comments on an earlier draft of this manuscript.

\end{document}